\documentclass[letter,scriptaddress,twocolumn, prd,showkeys]{revtex4}

	\usepackage{amsmath}%,amssymb}
	\usepackage{makeidx}
	\usepackage{amsfonts}
	\usepackage[ansinew]{inputenc}
	\usepackage[usenames,dvipsnames]{pstricks}
	\usepackage{subfigure}
	\usepackage{epsfig}
	\usepackage{pst-grad} % For gradients
	\usepackage{pst-plot} % For axes
	\usepackage[colorlinks,hyperindex]{hyperref}
	\hypersetup
	{
		colorlinks,%
		citecolor=black,%
		linkcolor=black,%
		urlcolor=black,%
	}

\makeindex

\begin{document}

\title{Strategy to Construct Exact Solutions in Einstein-Scalar Gravities }

\author{Qiang Wen}
\email{wen-q@ruc.edu.cn}
%\homepage{http://stoa.usp.br/thschiavo}
\affiliation{Department of Physics, Renmin University of China, Beijing 100872, China}

\affiliation{Perimeter Institute for Theoretical Physics, Waterloo, Ontario N2L 2Y5, Canada}

\affiliation{Yau Mathematical Sciences Center, Tsinghua University, Beijing 100084, China}

\date{\today}

\begin{abstract}
  We propose a new efficient strategy to construct exact solutions of Einstein-scalar gravities. We find that with some given metric $Ansatz$ the EOMs (equations of motion) are invariant under a rescaling operation along the radial direction, which makes the function of the scalar field scale invariant. Our strategy is to use the symmetry of the EOMs to give a scale invariant $Ansatz$ for scalar field first, then derive the metric and the corresponding scalar potential later. We construct large classes of exact solutions with two kinds of spherical metric, which include many new scalar hairy black holes  in different dimensions and some three-dimensional solitons and conical defects. We also discuss the thermodynamics of these solutions in general with Wald's formula.
\end{abstract}

\keywords{Einstein-scalar gravity, minimal coupled, exact solutions, scale invariant EOMs}

\maketitle
\tableofcontents

%\section{Introduction}
\section{Introduction}
Since the discovery of Einstein's General Relativity, there have been continuing efforts in constructing new exact solutions. In high energy physics, theories with gravity coupled to matter fields are studied frequently, string theory contains gauge fields and scalar fields, and its low energy effective theory, gauged supergravity, can be consistently truncated to gravity with a single scalar field~\cite{Duff:1999gh,Freedman:1999gk}, which is just a Einstein-scalar gravity with a super-scalar-potential. Setting $16\pi G=1$, the Lagrangian of a general d-dimensional Einstein-scalar gravity with a minimally coupled scalar field is given by
\begin{equation}\label{Lag}
\mathcal{L}=\sqrt{-g}\left(R-\frac{1}{2}(\partial \phi)^2-V(\phi)\right),
\end{equation}
the scalar potential $V(\phi)$ is a general function of $\phi$ with a Taylor expansion
\begin{align}\label{sp}
V(\phi)=-(d-1)(d-2)g^2+\frac{1}{2}m^2\phi^2+\gamma_3\phi^3+\cdots
\end{align}
where $d$ is the spacetime dimension, $m$ is the mass of the scalar field and the constant $g$ (like a gauge-coupling constant in gauged supergravity) can be taken as the inverse of the AdS radius $g=\frac{1}{\ell}$ if $g^2>0$. It is long been known that scalars are stable in $d+1$ dimensional AdS spacetime provided the mass $m^2$ is above the Breitenlohner-Freedman (BF)~\cite{BF} bound $m^2_{BF}=-\frac{d^2}{4\ell^2}$. Although we write $g^2$ as a square, it can be positive, zero and negative, so giving rise to solutions that are asymptotic anti-de-Sitter, flat or de-Sitter, respectively.

For a general metric, the EOMs contain the Einstein equations
\begin{align}
Z_{\mu\nu}=R_{\mu\nu}-\frac{1}{2}\partial_{\mu}\phi\partial_{\nu}\phi-\frac{1}{d-2}V(\phi)g_{\mu\nu}=0,
\end{align}
 and also the equation of motion for the scalar field,
\begin{align}\label{seom}
\nabla_{\mu}\nabla^{\mu}\phi=\frac{\partial V(\phi)}{\partial\phi},
\end{align}
where $\nabla_{\mu}$ are the covariant derivatives.

Exact solutions of such Einstein-scalar gravities are not only meaningful by themselves but also very useful in fields like AdS/CFT~\cite{Maldacena:1997re} correspondence, cosmology, black hole thermodynamics~\cite{Lu:2014maa,Wen:2015uma} and phase transitions~\cite{Hawking:1982dh}. The solutions could be even more attractive if the Einstein-scalar gravity is a truncation from string theory. In this article we concentrate on constructing exact solutions of these Einstein-scalar gravities (\ref{Lag}).

The first class of scalar hairy exact solutions are the 3-dimensional HMTZ (Henneaux, Martinez, Troncoso and Zanelli) black holes~\cite{Henneaux:2002wm}. However, as there is too much freedom in constructing a scalar potential without breaking any essential symmetry, starting from some arbitrary scalar potential, the possibility to find an exact solution could be almost null. Hence it is not surprising that there is not much progress~\cite{Martinez:2004nb,Kolyvaris:2009pc} in constructing exact solutions for a long time until recently~\cite{Anabalon:2012ta,Anabalon:2013sra,Anabalon:2013eaa,Anabalon:2013qua,Anabalon:2012dw,Acena:2012mr,Acena:2013jya,Gonzalez:2013aca,Correa:2011dt,Feng:2013tza,Faedo:2015jqa,Fan:2015tua,Li:2011hp,Cai:2012xh,Fan:2015oca}, people begin to think in a reverse way, which is trying to give a proper $Ansatz$ for the scalar field first then deriving the corresponding Lagrangian (or scalar potential) through the EOMs at last. 

However, starting from an arbitrary scalar field and metric $Ansatz$, it is still very difficult to find exact solutions. So it is important to find a way to choose promising $Ansatz$ for scalar field and metric, which is the central topic of this paper. In~\cite{Feng:2013tza}, we gave a proper $Ansatz$ for scalar field and metric with the inspiration from the construction of p-branes. 

In this paper, we go a step forward by using the symmetry of the EOMs to find some properties of the scalar field, thus giving some indications on how to choose a proper $Ansatz$ for the scalar field as well as the metric. In Sec. \ref{s2} we explicitly explain the spirit of our strategy by giving examples of constructing new exact solutions with Schwarzschild spherical metric (\ref{M1}), also we discuss the thermodynamics of these solutions and find some of these solutions have a manifest scalar charge (information about scalar charge can be found in \cite{Liu:2013gja,Lu:2014maa,Wen:2015uma}). To show the efficiency of our strategy, we give further examples in Sec. \ref{s3} to construct exact solutions with another spherical metric with a conformal factor (\ref{M2}). We give a discussion in the last section.

Through out this paper (except Sec. \ref{s34}), the parameters $\alpha$ and $g^2$, which we will encounter frequently, are integration constants which come out when we solve the EOMs. Since they appear in the scalar potential, they cannot be considered as charges (or physical integration constants) of the solutions. All our solutions (except the solutions presented in \ref{s34}) have only one physical integration constant $q$, which we take as a scalar charge.

\section{Constructing exact solutions with Schwarzschild spherical metric}\label{s2}
\begin{widetext}
\begin{align}
\label{eq1}
&\chi'+\frac{r\phi'^2}{2(d-2)}=0\,,
\\
\label{eq2}
&-\frac{6-2d}{r^2f}+\frac{6-2d}{r^2}+\frac{(6-2d)\chi'}{r}-\frac{(4-d)f'}{rf}
-\frac{3f'\chi'}{f}+2\chi'^2+\frac{f''}{f}-2\chi''=0\,,
\\
\label{eq3}
&\frac{6-2d}{r^2}-\frac{(6-2d)f}{r^2}+\frac{d~f'}{r}+\frac{2(1-d)f\chi'}{r}-3f'\chi'+2f\chi'^2+f''-2f\chi''+\frac{4V(\phi)}{d-2}=0.
\end{align}
\end{widetext}
\subsection{The strategy}
First we would like to look for static spherical solutions with Schwarzschild spherical metric $Ansatz$

\begin{align}\label{M1}
ds^2&=-h(r)dt^2+\frac{dr^2}{f(r)}+r^2 d\Omega_{d-2}^{2}\,,
\\
\phi&=\phi(r)\,,
\end{align}
where $d\Omega_{d-2}^{2}$ is the metric on the unit $(d-2)$-sphere. Define $h(r)=f(r)e^{-2\chi(r)}$ we can write the Einstein EOMs as (\ref{eq1})-(\ref{eq3}). The scalar field equation of motion (\ref{seom}) can be written as
\begin{align}\label{seom1}
 \frac{(d-2)f\,\phi'}{r}+\frac{1}{2}f'\,\phi'+\frac{f\, h' \,\phi'}{2 h}+f\, \phi''=\frac{\partial V(\phi)}{\partial\phi}\,.
\end{align}

In order to make this strategy work out well, we have reorganised the EOMs to be in a proper form, so we can derive them one by one. For example in (\ref{eq1}) we can derive $\chi(r)$ from the given $Ansatz$ for the scalar field, then derive $f(r)$ from (\ref{eq2}), and finally get the scalar potential $V(\phi)$ from (\ref{eq3}). These three Einstein equations (\ref{eq1})-(\ref{eq3}) are enough to derive all the unknown metric functions and the corresponding scalar potential, one can check that, (\ref{seom1}) is automatically satisfied by all our solutions in this section.

We can see that when $d=3$ the Einstein equations (\ref{eq1})-(\ref{eq3}) is invariant under a rescaling operation
\begin{align}\label{res1}
&r=c\rho\qquad& \phi(c\rho)=\tilde{\phi}(\rho)\,,
\cr
&f(c\rho)=c^2\tilde{f}(\rho)\qquad &~\chi(c\rho)=\tilde{\chi}(\rho)\,,
\cr
&V(\phi)=\tilde{V}(\tilde{\phi})\,.  \qquad &
\end{align}
Substitute (\ref{res1}) into (\ref{eq1})-(\ref{eq3}), the Einstein equations are slightly modified with the constant $6-2d$ in the first terms of (\ref{eq2}) and (\ref{eq3}) rescaled to $(6-2d)/c^2$ (leaving the other coefficients unchanged). One may notice immediately that when $d=3$, this constant vanishes, so the EOMs are totally invariant under (\ref{res1}). This means $\tilde{\phi}(\rho),\tilde{\chi}(\rho),\tilde{f}(\rho)$ are still solutions of (\ref{eq1})-(\ref{eq3}) when $d=3$, thus the function $\phi(r)$ should be preserved by the rescaling operation (\ref{res1}). From now on we concentrate on three-dimensional theories in this section.

We can write the asymptotic behavior of $\phi(r)$ in the following form
\begin{equation}\label{lre}
\phi(r)=\sum_{i=1}^{\infty}\frac{\phi_i}{r^{p_i}}\,,
\end{equation}
where the exponents $p_i$ have a relationship with dimension $d$ and the scalar field mass $m$ which can be read from the scalar potential (\ref{sp}). For example, define
\begin{align}
\sigma=\sqrt{4l^2m^2+(d-1)^2}\,, 
\end{align} when $0<\sigma<1$, by solving the linearized equation of motion for the scalar field $(\nabla_{\mu}\nabla^{\mu}-m^2)\phi=0$, we will get
\begin{equation}\label{pi}
 p_1=\frac{d-1-\sigma}{2}\,, \qquad p_2=\frac{d-1+\sigma}{2}\,.
 \end{equation}
So if we give the scalar field first, the scalar potential we derived should give the right $m$ which satisfies (\ref{pi}).

The static solutions have at most two physical integration constants (constants which only appear in the solution, not in the Lagrangian, we can take these constants as charges of the solutions) for a given theory, for example we can take them as the radius of the event horizon $r_0$ and $\phi_1$. We can see this from the numerical studies in~\cite{Wen:2015uma} and a class of new exact solutions in Sec. \ref{s34}. So $\phi_i$ should be a function of $r_0$ and $\phi_1$. Since the function of our solution $\phi(r,r_0,\phi_1)$ is preserved by the rescaling (\ref{res1}), we find the coefficients $\phi_i$ in (\ref{lre}) with $i\geq 2$ should look like
\begin{equation}\label{phii}
\phi_i(r_0,\phi_1)=\sum_{n}\mathcal{D}_n\frac{\phi_1^{(n+p_i)/p_1}}{r_0^{n}}\,,
\end{equation}
where $\mathcal{D}_n$ are coefficients independent of $r_0$ and $\phi_1$. Thus the rescaling operation changes only the integration constants $r_0$ and $\phi_1$, while the function of the scalar field itself is unchanged:
\begin{align}
\phi(r,\phi_1,r_0)~\rightarrow~\phi(\rho,\phi_1',r_0')\,.
\end{align}

Now we know that our $Ansatz$ for the scalar field should satisfy (\ref{lre}) and ({\ref{phii}). If we keep all the terms in the summation in (\ref{phii}), (\ref{lre}) could be a complicated $Ansatz$ for the scalar field. However, for every $\phi_i$, there is always a $r_0$ independent term which correspond to the $n=0$ term in ({\ref{phii}), if we only keep this term for every $\phi_i$, we can give a much simpler $Ansatz$ for the scalar field. The scalar field chosen by this way have an asymptotic behavior as follows
\begin{equation}\label{exp}
\phi(r)=\sum_{i=1}^{\infty}\mathcal{C}_i\frac{\phi_1^{p_i/p_1}}{r^{p_i}}\,.
\end{equation}
Here we drop the other physical integration constant $r_0$ and the boundary condition $\phi_2\sim \phi_1^{p_2/p_1}$ is automatically given, so starting from an $Ansatz$ satisfying (\ref{exp}) we may get solutions with only one physical integration constant (at least in the scalar field), which we will see later, the only found solution with two physical integration constant is given in the subsection \ref{s34}. We can further simplify the scalar field $Ansatz$ by just choosing $\phi(r)$ as simple analytical functions of $\frac{\phi_1}{r^{p_1}}$ which will absolutely satisfy (\ref{exp}).

To make our strategy more clear, we conclude its application with the following steps:

1) Choose a proper metric $Ansatz$ so that the EOMs have a scale invariance\,,

2) Give an $Ansatz$ for the scalar field which satisfies (\ref{lre}) and (\ref{phii})\,,

3) Substitute the scalar field $Ansatz$ into the EOMs, then derive the metric functions and the corresponding scalar potential.

Since the simpler $Ansatz$ we give for the scalar field the bigger possibility we will have to find exact solutions, we choose simple analytic functions of $\frac{\phi_1}{r^{p_1}}$ as $Ansatz$es for the scalar field. To find soliton solutions, which have no event horizon and singularity, the $Ansatz$ for the scalar field should be regular everywhere and must satisfy (\ref{exp}), because static solitons only have one integration constant $\phi_1$\,.

It is true that our strategy is still a method about trying, however, unlike the scalar potential which we have great freedom to construct, there are not many simple analytic functions for $\phi(r)$ to choose. And absolutely we cannot guarantee to find all exact solutions, but if there exist simple exact solutions it would be quite promising to find them with our strategy. Our strategy only works in three-dimensions for the Schwarzschild spherical static metric (\ref{M1}), because the scale invariance of the EOMs are broken in higher dimensions thus will not admit simple analytic functions of $\frac{\phi_1}{r^{p_1}}$ as a solution of $\phi(r)$. This explains why there is still no exact solutions found in higher dimensions (numerical solutions in higher dimensions are studied in~\cite{Lu:2014maa,Hertog:2004dr,Hertog:2004ns,Hertog:2005hu,Craps:2007ch}).

\subsection{The thermodynamics}

Before we use our strategy to construct exact solutions in the following subsections, we use Wald's formula~\cite{Wald:1993nt,Iyer:1994ys} to calculate their thermodynamic first law in general. This is to construct a closed $(d-2)$-form $(\delta Q-i_{\xi}\Theta)$ where $\xi$ is the Killing vector. Taking $\xi=\partial/\partial t$ and applying this to our Lagrangian and metric $Ansatz$, Wald's formula states that when the metric and matter fields are on shell, the integral of $(\delta Q-i_{\xi}\Theta)$ over any $S^{d-2}$ surface at constant $t$ and radius $r$ is independent of $r$. The $(d-2)$-form $\xi\cdot\Theta$ has contributions from both the gravity sector $\xi\cdot\Theta^{G}$ and the scalar field sector $\xi\cdot\Theta^{\phi}$, after some calculation we get
\begin{subequations}
\label{Theta}
\begin{align}
&\xi\cdot\Theta^{G}=\epsilon_{i_1\cdots i_{d-2} t \mu}(g^{\mu m}g^{\nu n}-g^{\mu n}g^{\nu m})\nabla_{n}\delta g_{\nu m},
\\
&\xi\cdot\Theta^{\phi}=-\epsilon_{i_1\cdots i_{d-2} t }~^{\mu}(\nabla_{\mu}\phi~\delta \phi),
\\
&Q=\epsilon_{i_1\cdots i_{d-2}}~^{\mu t}\nabla_{\mu}\xi_{t},
\end{align}
\end{subequations}
where $\epsilon$ is the Levi-Civita tensor. 

It is convenient to define
\begin{align}
\delta\mathcal{H}_r=\int_{S_r}\,(\delta Q-i_{\xi}\Theta)\,,
\end{align}
which is the integral on a (d-2)-sphere with a radius $r$.
For black holes we take one integral surface as the event horizon $S_{r_0}$ while the other as the infinite far away boundary $S_\infty$, so Wald's formula gives
\begin{align}\label{wfb}
\delta\mathcal{H}_{r_0}=\delta\mathcal{H}_\infty\,.
\end{align}
For solitons (or bare singularities) we take one integral surface as the infinite small circle surrounded the origin $S_{0^+}$ while the other as $S_\infty$, hence Wald's formula would give
\begin{align}\label{wfs}
\delta\mathcal{H}_{0^{+}}=\delta\mathcal{H}_\infty\,.
\end{align}

For the metric (\ref{M1}) we choose in this section, this integral has already been calculated in~\cite{Liu:2013gja,Lu:2014maa}, for an arbitrary sphere $S_r$ this integral can be expressed as
\begin{align}\label{fl1}
\delta\mathcal{H}_r
=-\omega_{d-2}r^{d-2}\sqrt{\frac{h}{f}}\left(\frac{d-2}{r}\delta f+f\phi'\delta\phi\right)
\end{align}
where $w_{d-2}$ is the volume of the unit $S^{d-2}$ sphere. Substitute this and the metric functions into (\ref{wfb}) or (\ref{wfs}), we would get the thermodynamic first law of the solutions.

It can be seen that for solitons in $d\geq 4$ dimensions, $\delta\mathcal{H}_{0^{+}}$ would vanish, however, as we will see, we can get nonzero $\delta\mathcal{H}_{0^{+}}$ in three-dimensions. This part of contribution in the first law should not be considered as a contribution from $T\delta S$ since there is no horizon and the temperature is not well defined, we would consider this term as a contribution from the scalar charge $\Phi_S\delta Q_S$.

For black holes, it has already been shown in~\cite{Lu:2014maa,Liu:2013gja} that the left-hand side of Wald's formula (\ref{wfb}) is given by
\begin{equation}
\delta\mathcal{H}_{r_0}
=\frac{(d-2)\omega_{d-2}}{16\pi G}\sqrt{f'(r_0)h'(r_0)} r_0^{d-3}\,\delta r_0=T\,\delta S
\end{equation}
where the Hawking temperature $T$ and entropy $S$ are given by
\begin{align}
T=\frac{\sqrt{f'(r_0)h'(r_0)}}{4\pi}\,,\qquad S=\frac{A}{4 G}=\frac{r_0^{d-2}\omega_{d-2}}{4G}\,.
\end{align}
The right-hand side of (\ref{wfb}) is defined as the variation of the mass $\delta M_W$ (we call the mass defined by Wald's formula as $M_W$), thus Wald's formula just gives the standard thermodynamic first law $\delta M_W=T\,\delta S$.

It should be noted that the boundary conditions $\phi_2(\phi_1)$ of our scalar field $Ansatz$es given by (\ref{exp}) preserve all the asymptotic AdS symmetries. The analysis of~\cite{Wen:2015uma,Lu:2014maa,Anabalon:2014fla} showed that, under such special boundary conditions, the masses calculated by Wald's formula~\cite{Wald:1993nt,Iyer:1994ys}, the Hamiltonian formula~\cite{Hertog:2004dr,Henneaux:2006hk} and the AMD~\cite{Ashtekar:1984zz,Ashtekar:1999jx} conformal method are the same. 

According to~\cite{Wen:2015uma}, in Einstein-scalar gravities, the masses calculated by Wald's formula and the Hamiltonian formula are defined with a boundary condition. In fact there is no restriction for choosing a boundary condition in the space of solutions. People usually prefer to choose boundary conditions which preserve all the asymptotic AdS symmetries, under which the first law is just given by $\delta M_{W}=T\delta S$. If we consider holography~\cite{Maldacena:1997re}, we need a boundary condition since it determines the action of the dual quantum field theory~\cite{Witten:2001ua,Berkooz:2002ug,Sever:2002fk}, the so-called ``Designer gravity"~\cite{Hertog:2004ns} is a method to construct different boundary quantum field theories by choosing different boundary conditions on the dual Einstein-scalar gravity side. However, if we also define the Einstein-scalar gravity with a boundary condition, we may have problems~\cite{Wen:2015uma}.

A new boundary condition independent definition of mass is given in~\cite{Wen:2015uma}, which is similar in spirit to both Wald's formula and the Hamiltonian formula, with the only difference that we require the variation of the mass to have no contribution from the variation of the matter charges. In this paper we will use this definition to calculate the mass, and use Wald's formula to derive the thermodynamic first law. 

For self-consistency, we give a brief review on this definition of mass here. The mass have the same formula as Wald's definition~\cite{Wald:1993nt,Iyer:1994ys}, which is the surface integral of the closed $(d-2)$-form on the infinite faraway boundary
\begin{equation}\label{Nm}
\bar{\delta}M=\int_{S_\infty}(\bar{\delta} Q_{\xi}-\xi\cdot\Theta(\varphi,\bar{\delta}\varphi))\,,
\end{equation}
with the only difference that we use a variation $\bar{\delta}$ which only act along the mass direction while the variation $\delta$ in Wald's definition act along all independent charges. This means for any charge $Q$ we should have $\bar{\delta}Q=0$ unless $Q$ is the mass $M$. Arguments for why we define the mass in this way can be found in~\cite{Wen:2015uma}.

In our cases we have two charges, which are the mass and the scalar charge. It should be noted that, although the scalar charge arises as an integration constant in the solutions, there is no symmetry correspond to this charge~\cite{Gibbons:1996af}. Which means, unlike the mass, the scalar charge is not conserved and we cannot calculate it the way we calculate a Noether charge. Usually we take the integration constant $\phi_1$ as the scalar charge~\cite{Lu:2014maa,Wen:2015uma,Liu:2013gja,Gibbons:1996af}
\begin{align}
Q_S=\phi_1.
\end{align}

To integrate out $M$, first we substitute the asymptotics of the metric functions and scalar field into (\ref{Nm}), take the limit $r\to\infty$, then with a Legendre transformation, we write the right-hand side as a total variation plus a scalar charge term $\Phi_S\bar{\delta}\phi_1$ (where $\Phi_S$ is defined as the conjugate potential of $Q_S$), then at last we impose $\bar{\delta}\phi_1=0$, hence, the integration of the total variation would give our mass.

For our cases in this section, the asymptotics of the metric functions and scalar field will take the form~\cite{Lu:2014maa}
\begin{subequations}\label{sle}
\begin{align}
\phi &= \frac{\phi_1}{r^{(d-1-\sigma)/2}}+\frac{\phi_2}{r^{(d-1+\sigma)/2}}
 + \cdots \,,
\\
h&= \frac{r^2}{\ell^2} + 1-\delta_{d,3} + \frac{\kappa}{r^{d-3}} +
\cdots\,,
\\
f &=\frac{r^2}{\ell^2} + 1-\delta_{d,3} + \frac{b}{r^{d-3-\sigma}} + \frac{\beta}{r^{d-3}} +
\cdots.
\end{align}
\end{subequations}
\begin{widetext}
Substituting the expansions (\ref{sle}) into the equations of
motion and solving for the first few coefficients, we get
\begin{subequations}
\begin{align}
&b=\frac{(d-1-\sigma)\, \phi_1^2}{4(d-2)\ell^2}\,,
\\
&\beta=\kappa + \frac{[(d-1)^2-\sigma^2]\, \phi_1\phi_2}{2(d-1)(d-2)\ell^2}\,.
\end{align}
\end{subequations}

We consider the definition of mass (\ref{Nm}) and get
\begin{align}
\bar{\delta}M =& \omega_{d-2}\Big[
  -(d-2)\bar{\delta}\kappa +
\frac{\sigma}{2(d-1)\ell^2}\,
   [(d-1+\sigma)\phi_2\bar{\delta}\phi_1 - (d-1-\sigma)\phi_1\bar{\delta}\phi_2]\Big]
   \nonumber\\
=& \omega_{d-2}\Big[\bar{\delta}\left( -(d-2)\kappa - \frac{\sigma(d-1-\sigma)}{2(d-1)\ell^2}\phi_1\phi_2\right)+
\frac{\sigma}{\ell^2}\,
  \phi_2\bar{\delta}\phi_1\Big]
\nonumber\\
=&\bar{\delta}\left[\omega_{d-2}\left( -(d-2)\kappa - \frac{\sigma(d-1-\sigma)}{2(d-1)\ell^2}\phi_1\phi_2\right)\right]\,,
\label{deltaHgen}
\end{align}
\end{widetext}
where we have imposed $\bar{\delta}\phi_1=0$ in the third line of the equation, and $\omega_{d-2}$ is the volume of the unit $(d-2)$-sphere.
So the mass $M$ is given by
\begin{equation}\label{Mn1}
M=\omega_{d-2}\left( -(d-2)\kappa - \frac{\sigma(d-1-\sigma)}{2(d-1)\ell^2}\phi_1\phi_2\right)\,.
\end{equation}

We can also find
\begin{align}
\delta\mathcal{H}_{\infty}=\delta M+\frac{\omega_{d-2}\sigma}{\ell^2}\,\phi_2\delta\phi_1=\delta M-\Phi_S\delta Q_S
\end{align}
where the coefficient $\Phi_S=-\frac{\omega_{d-2}\sigma}{\ell^2}\,\phi_2$ is defined as the conjugate potential of the scalar charge. The first law calculated by Wald's formula is then given by
\begin{align}\label{FL2}
T\delta S=\delta M-\Phi_S\,\delta Q_S\,.
\end{align}

For our following 3-dimensional exact solutions constructed in this section,  the mass $M$ is given by
\begin{equation}\label{Mn}
M=\omega_1\left( -\kappa - \frac{\sigma(2-\sigma)}{4\ell^2}\phi_1\phi_2\right)\,,
\end{equation}
where $\omega_1=2\pi$, and the scalar potential is given by
\begin{align}\label{Scharge}
\Phi_S=-\frac{\omega_1\,\sigma}{\ell^2}\phi_2\,. 
\end{align}

\subsection{3-d solutions}
\subsubsection{\bf{Black holes with the simplest scalar field}}

We first consider the simplest $Ansatz$ for $\phi(r)$
\begin{align}\label{as1}
\phi(r)= \left(\frac{q}{r}\right)^{\mu}\,,
 \end{align}
where $q$ is the parameter which describes the scalar hair and $\mu$ is a positive number. The solutions could be complex for a general $\mu$, so we only consider the $\mu=1$ and $\frac{1}{2}$ cases, which will produce simple metric solutions and $q$ would not appear in the scalar potential $V(\phi)$, thus can be considered as a charge.

 When $\mu={\bf 1}$, the metric functions are given by
\begin{align}\label{mmu1}
h(r)&=\left(\alpha (e^{-\frac{q^2}{4 r^2}}-1)+g^2\right) r^2\,,
\cr
f(r)&=e^{\frac{q^2}{4r^2}}\left(\alpha+e^{\frac{q^2}{4r^2}}(g^2-\alpha)\right)r^2\,.
\end{align}
Both of these two metric functions go to $g^2 r^2$ asymptotically, so we can interpret $g$ as the inverse of the AdS radius, $g=1/\ell$. For a positive (zero or negative) $g^2$ we correspondingly get asymptotic AdS (flat or dS) solutions respectively. It is obvious that when the scalar hair vanishes, we get the massless static BTZ~\cite{Banados:1992wn} solution. The corresponding theory that admit solutions (\ref{as1}) and (\ref{mmu1}) is marked by the following scalar potentials
\begin{align}\label{spmu1}
V(\phi)=-2\alpha\, e^{\frac{\phi^2}{4}}-\frac{1}{2}\,e^{\frac{\phi^2}{2}}\,(\alpha-g^2)(\phi^2-4)\,.
\end{align}

The solution of $h(r)=0$ gives the radius of the event horizon of a black hole
\begin{align}
 r_0=\frac{q}{2\sqrt{\ln\frac{\alpha}{\alpha-g^2}}}\,.
 \end{align}
Thus for theories with a scalar potential (\ref{spmu1}), the condition for the existence of black hole solutions described by (\ref{mmu1}) is just
\begin{align}
 \alpha\,>\,g^2\,.
 \end{align}
 
According to (\ref{mmu1}) we have
\begin{align}
\kappa=-\frac{\alpha q^2}{4}\,,\qquad \phi_1=q\,, \qquad \phi_2=0\,,
\end{align}
so we have
\begin{align}
M=\frac{\pi\alpha\,q^2}{2}\,, \qquad\Phi_S=0\,,
\end{align}
where we can see the conjugate potential of the scalar charge is zero. Thus there is no contribution from the scalar charge in the first law. On the other hand we have
\begin{align}
 \delta\mathcal{H}_{r_0}=\pi\alpha\,q\delta q=T\delta S\,,
 \end{align} 
thus the thermodynamic first law is just
\begin{align}
\delta M=T\delta S\,.
\end{align}

When $\mu={\bf\frac{1}{2}}$ we have
\begin{align}\label{mmu2}
h(r)&=r^2\,\left(g^2-\alpha+\alpha\, (1+\frac{q}{8r})\,e^{-\frac{q}{8r}}\right)\,,
\cr
f(r)&=e^{\frac{q}{8r}}\,r^2\,\left((g^2-\alpha)e^{\frac{q}{8r}}+\alpha\,(1+\frac{q}{8r})\right)\,,
\end{align}
and both of the metric functions go to $g^2 r^2$ asymptotically.

The equation $h(r_0)=0$ has a root at
\begin{align}
r_0=-\frac{q}{8\,(1+ProductLog(\frac{g^2-\alpha}{e\,\alpha}))}
\end{align}
(where function $ProductLog(z)$ is defined as the solution of $x\,e^x=z$).
One can check that the condition for the existence of black hole solutions is also
\begin{align}
 \alpha\,>\,g^2\,.
\end{align}

The corresponding scalar potential in this case is given by
\begin{align}
V(\phi)=\frac{e^{\phi^2/8}}{8}\,\left(e^{\phi^2/8}(g^2-\alpha)\,(\phi^2-16)-\alpha\,(\phi^2+16)\right)
\end{align}

According to (\ref{mmu2}) we have
\begin{align}
\kappa=-\frac{\alpha\, q^2}{128}\,,\qquad \phi_1=\sqrt{q}\,, \qquad \phi_2=0\,,
\end{align}
so
\begin{align}
M=\frac{\pi\alpha\,q^2}{32}\,, \qquad\Phi_S=0\,,
\end{align}
which means the conjugate potential of the scalar charge is also zero.
We can calculate out
\begin{align}
\delta\mathcal{H}_{r_0}=\frac{\pi\alpha}{16}\,q\delta q=T\delta S\,,
\end{align} 
thus the thermodynamic first law is again
\begin{align}
\delta M=T\delta S\,.
\end{align}

\subsubsection{\bf{Black holes with scalar field described
by Arctanh functions}}

The second scalar field $Ansatz$ we try is 
\begin{equation}\label{sa2}
\phi(r)=2\sqrt{2\mu}~\text{Arctanh}\frac{1}{\sqrt{1+r/q}}\,,
\end{equation}
whose asymptotic behavior satisfies (\ref{exp}) and is given by
\begin{equation}
\phi(r)=2\sqrt{2\mu}(\frac{q}{r})^{1/2}-\frac{\sqrt{2\mu}}{3}(\frac{q}{r})^{3/2}+\cdots\,.
\end{equation}
We can read
\begin{align}
 p_1&=\frac{1}{2}\,,\qquad ~~~~~~~p_2=\frac{3}{2}\,,\qquad \sigma=1\,,
 \\
 \phi_1&=2\sqrt{2\mu q}\,,\qquad \phi_2=-\frac{\sqrt{2\mu \,q^3}}{3}\,,
 \end{align}
which indicates there is a nontrivial contribution from the scalar charge $\Phi_S\delta Q_S$ with
\begin{align}
Q_S=2\sqrt{2\mu q}\,,\qquad \Phi_S=\frac{2\pi\sqrt{2\mu \,q^3}}{3\ell^2}\,,
\end{align}
in the thermodynamic first law.

We substitute this scalar field into the EOMs and define $H=1+q/r$ to get
\begin{align}
\label{fr}
&f(r)=r^2 H^{\mu}\left(H^{\mu}(g^2-\alpha)+H(H+(\mu-2)\frac{q}{r})\alpha\right)\,,
\\
&h(r)=f(r)H^{-2\mu}\,.
\end{align}
It can be checked that the scalar equation of motion (\ref{seom1}) is also satisfied by these metric functions. The most general corresponding scalar potential is given by (\ref{V2}),
where $\varphi=\frac{\phi}{2\sqrt{2\mu}}$. The expansion of $V$ around $\phi=0$ is
\begin{align}
V(\phi)=-2g^2-\frac{3g^2\phi^2}{8}-\frac{g^2\phi^4}{32}+\cdots\,,
\end{align}
where we can read the mass of the scalar field $m^2=-\frac{3g^2}{4}$, which is above the Breitenlohner-Freedman (BF) bound~\cite{BF} and indicates $p_1=\frac{1}{2}$ and $p_2=\frac{3}{2}$ as expected.

\begin{widetext}
\begin{equation}\label{V2}
V(\varphi)=(\text{cosh}~\varphi)^{2(\mu+1)}\left(\alpha((2-\mu)\tanh^2\varphi-2)+(g^2-\alpha)(\cosh\varphi)^{2(\mu-1)}(\mu\tanh^2\varphi-2)\right)
\end{equation}
\begin{align}\label{ceh'}
&h(r)=(\mu-1)q^{2-\mu}\alpha r^{\mu}+\mu(2-\mu) q^{1-\mu}\alpha r^{\mu+1}+(g^2-\alpha)r^2+\cdots
\\
\label{cef'}
&f(r)=(\mu-1)q^{2+\mu}\alpha r^{-\mu}+\mu^2q^{1+\mu}\alpha r^{1-\mu}+q^{2\mu}(g^2-\alpha)r^{2-2\mu}+\cdots
\end{align}
\end{widetext}

 Metric functions $h(r)$ and $f(r)$ all go to $g^2r^2$ asymptotically, and their Taylor expansions are (\ref{ceh'}) and (\ref{cef'}). If $h(0)<0$, the equation $h(r)=0$ have at least one root, which guarantees the existence of a black hole. This can be translated to the following conditions
\begin{subequations}\label{eh}
\begin{align}
&0<\mu<1,~ \alpha>0\,,
\\
& 1<\mu<2,~\alpha<0\,,
\\
&\mu>2,~ \alpha>g^2.
\end{align}
\end{subequations}

The asymptotic behavior of $h(r)$ is
\begin{align}
h(r)=g^2 r^2-\frac{1}{2}(2-3\mu+\mu^2)q^2\alpha+\cdots\,,
\end{align}
which indicates $\kappa=-\frac{1}{2}(2-3\mu+\mu^2)q^2\alpha$, so the mass is given by
\begin{align}\label{mass2}
M=\pi(2-3\mu+\mu^2)\,q^2\alpha+\frac{2\pi\mu}{3\ell^2}\,q^2.
\end{align}

Theories satisfy (\ref{eh}) admit black hole solutions, for example setting $\mu=3$, the scalar potential and solution become
\begin{align}
&V(\varphi)=-\frac{1}{2}\cosh^6\varphi~\{\alpha-(g^2-\alpha)\cosh^4\varphi(\cosh[2\varphi]-5)
\nonumber
\\
&~~~~~~~~~~+3\alpha\cosh[2\varphi]\}\,,
\\
&ds^2=-r^2\left(g^2-\frac{q^2\alpha}{(q+r)^2}\right) dt^2
\nonumber\\
&~~~~~~~~+\frac{r^4}{(q+r)^4(g^2(q+r)^2-q^2\alpha)}dr^2+r^2 d\theta^2.
\end{align}
There is a horizon at $r_0=q(\sqrt{\alpha/g^2}-1)$, where $\alpha>g^2$ is required by (\ref{eh}). The Hawking temperature $T$ and entropy $S$ are given by
\begin{align}
T=\frac{q\alpha}{2\pi (\sqrt{\alpha/g^2}-1)}\qquad S=\frac{\pi(\sqrt{\alpha/g^2}-1)q}{2}.
\end{align}
We can check the first law $\delta M=T \delta S+\Phi_S\delta Q_S$ is satisfied.

{\bf When ${\bf \mu=1,2}$}, the free parameter $\alpha$ disappears in the scalar potential and metric, in these special theories, the solutions can no longer be fully described by (\ref{fr}) and (\ref{V2}). When $\mu=1$, the scalar potential and solution are
\begin{align}
&V(\varphi)=- g^2\cosh^4\varphi~\{2+4\alpha \ln(\text{cosh}\varphi)
\nonumber
\\
&~~~~~~~~~~+(2\alpha\ln(\text{sech}\varphi)-2\alpha-1)\tanh^2\varphi\},
\\
&h(r)=g^2 r^2\left(1-\frac{\alpha q}{r}-\alpha \ln\frac{r}{r+q}\right),
\\
&\chi(r)=\ln\frac{r+q}{r}\,,
\end{align}
and when $\mu=2$, we get
\begin{align}
&V(\varphi)= g^2\cosh^8\varphi~\{(4\alpha\ln(\cosh\varphi)-2)\text{sech}^2\varphi
\nonumber
\\
&~~~~~~~~~~-\alpha(1-\text{sech}^4\varphi)\}\,,
\\
&h(r)=g^2 r^2\left(\frac{\alpha q}{r+q}+1+\alpha\ln\frac{r}{r+q}\right),
\\
&\chi(r)=2\ln\frac{r+q}{r}\,.
\end{align}
This $\mu=2$ case reproduces a class of solutions reported in~\cite{Correa:2011dt}.

Theories with $\alpha=0$ have solutions
\begin{align}
&V(\varphi)=g^2 \cosh^{4\mu}\varphi(\mu\tanh^2\varphi-2)\,,
\\
&ds^2=-g^2 r^2 dt^2+\frac{1}{g^2 r^2(q/r+1)^{2\mu}} dr^2+r^2 d\theta^2\,,
\end{align}
with a nonzero mass which is related to the scalar hair
\begin{align}
M=\frac{2\pi\mu}{3\ell^2}\,q^2\,.
\end{align}
However, the masses defined by Wald's formalism, the Hamiltonian formula and the AMD conformal method are all zero and independent of the physical integration constant $q$, which will lead to the problem of how to explain the entropy of these black holes, for details, see~\cite{Wen:2015uma}.

As the scalar field (\ref{sa2}) always has a {\em{log}}(r) singularity on the origin, there is no soliton solution.

We find the HMTZ black holes are in fact contained in
our solutions. One of our principles for choosing scalar
field $Ansatze$ is "the simpler the better". However, the scalar
field of the HMTZ black holes seems not simple at all, so it
could be quite difficult to think of starting from such a
scalar field to construct exact solutions with our strategy. So
it is with a little surprise that we find the scalar field of the
HMTZ solutions is, in fact, a special case of our $Ansatz$ (\ref{sa2}). 

The scalar field of the HMTZ black holes~\cite{Henneaux:2002wm} is given by
\begin{equation}\label{as3}
\phi(r)=4~\text{Arctanh} \sqrt{\frac{B}{B+\frac{1}{2}(r+\sqrt{4Br+r^2})}}\,,
\end{equation}
(the additional factor 4 comes out because our Lagrangian is written in a different form) which looks quite different from our $Ansatz$ (\ref{sa2}). Define $x=\frac{r}{4B}$ and $q=4B$, then use the following equation
\begin{align}\label{equi}
\text{Arctanh} \frac{1}{\sqrt{1+x}}=2~\text{Arctanh} \frac{1}{\sqrt{1+2 x(1+\sqrt{1+1/x})}}\,,
\end{align}
we find scalar field (\ref{as3}) reduces directly to
\begin{align}
\phi(r)=2~\text{Arctanh} \frac{1}{\sqrt{1+r/q}}.
\end{align}
This scalar field is just our $Ansatz$ (\ref{sa2}) with $\mu=\frac{1}{2}$, which satisfies (\ref{eh}) and thus, would admit black hole solutions.
Furthermore, we redefine $\alpha=\frac{1}{2}g^2(1+\nu)$, and the scalar potential (\ref{V2}) reduces to
\begin{align}
V=-2g^2(\cosh^6[\frac{\phi}{4}]+\nu \sinh^6[\frac{\phi}{4}])\,,
\end{align}
which is exactly the scalar potential that gives the HMTZ black hole solutions. So the HMTZ black holes are just some special cases of the larger class of exact solutions we constructed with our strategy in this subsection.

\subsubsection{\bf{Solitons and black holes with scalar field described
by Arcsin functions}}

In this subsection we try to find exact soliton solutions with a scalar field $Ansatz$
given by
\begin{equation}\label{sa3}
\phi(r)=4 \mu ~\text{ArcSin} \sqrt{\frac{1}{1+r/q}}\,,
\end{equation}
where $q$ and $\mu$ are non-negative constants. These scalar fields are regular everywhere and their asymptotic behavior
\begin{align}
 \phi(r)=4\mu(\frac{q}{r})^{\frac{1}{2}}-\frac{4\mu}{3}(\frac{q}{r})^{\frac{3}{2}}+\cdots\,,
 \end{align} 
satisfies (\ref{exp}) and gives
\begin{align}
p_1&=\frac{1}{2}\,,\qquad ~~~~~p_2=\frac{3}{2}\,,\qquad \sigma=1\,,
 \\
\phi_1&=4\mu \,q^{\frac{1}{2}}\,,\qquad \phi_2=-\frac{4\mu}{3}\,q^{\frac{3}{2}}\,.
 \end{align}
which means
\begin{align}\label{Phi}
Q=4\mu \,q^{\frac{1}{2}}\,,\qquad \Phi=\frac{8\pi\mu}{3\ell^2}\,q^{\frac{3}{2}}\,.
\end{align}
Since this $Ansatz$ is a multiple valued function, we confine $0<\frac{\phi(r)}{4\mu}\leq\frac{\pi}{2}$.

For a general $\mu$, the analytic solution for $f(r)$ is a little complex, so we get involved with exponential integral functions. Here we set $\mu=1$, and the resulting solutions are much simpler. Substitute our $Ansatz$ (\ref{sa3}) into the EOMs, and we get
\begin{subequations}\label{msoliton}
\begin{align}
h(r)&=(g^2-\alpha)r^2+\alpha e^{-\frac{2q}{r+q}}(r+q)^2\,,
\\
f(r)&=e^{\frac{4q}{r+q}}(g^2-\alpha)r^2+\alpha e^{\frac{2q}{r+q}}(r+q)^2\,,
\end{align}
\end{subequations}
where $g^2$ and $\alpha$ are constants. 
The corresponding scalar potential $V(\phi)$ is given by
\begin{align}\label{sp3}
&V(\phi)=
\nonumber\\
&\frac{1}{4}e^{4\sin^2\frac{\phi}{4}}(\alpha-g^2)(7+\cos \phi)-2\alpha e^{2\sin^2\frac{\phi}{4}}\sec^2\frac{\phi}{4}\,. 
\end{align}
Again, as expected, Eq. (\ref{seom1}) is satisfied.

The Taylor expansion of $V(\phi)$ is given by
\begin{align}
&V(\phi)=-2g^2-\frac{3g^2\phi^2}{8}-\frac{g^2\phi^4}{32}+\cdots\,,
\end{align}
from which we can read $m^2=-\frac{3}{4}g^2$, which is above the BF bound~\cite{BF} and gives $p_1=\frac{1}{2},~p_2=\frac{3}{2}$, as expected.

\begin{figure}

\centering
\includegraphics[width=0.4 \textwidth]{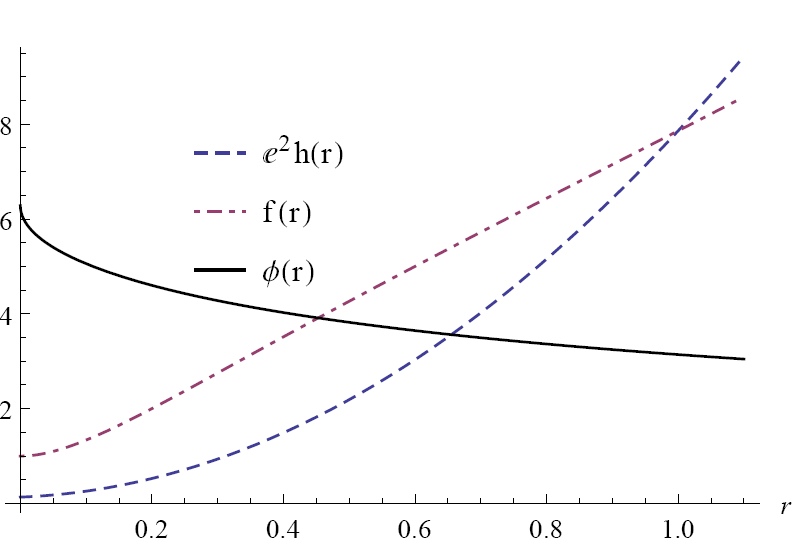}
\hfill
\includegraphics[width=0.4 \textwidth]{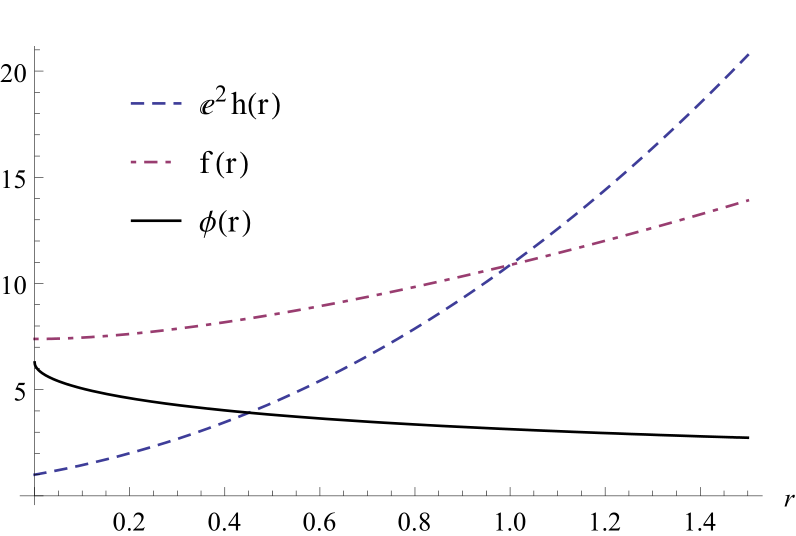}
\hfill
\includegraphics[width=0.4 \textwidth]{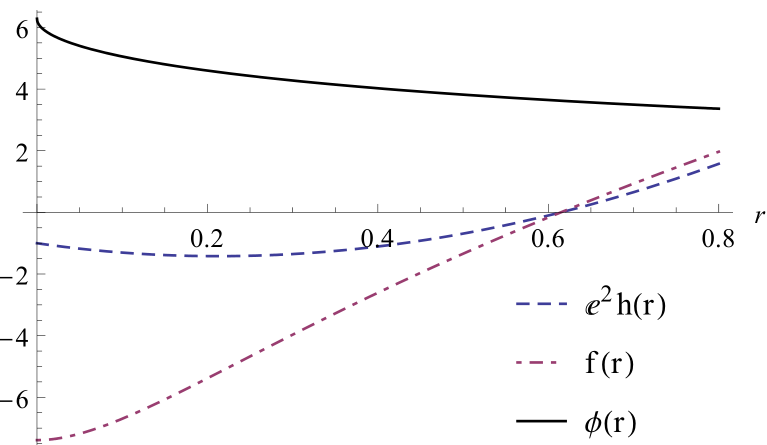}
\caption{\label{f1} Setting $g^2=q=1$ and with $\alpha=1/e^2,~1$ and $-1$ respectively, from top to bottom this figure shows the scalar field and metric functions of a soliton, conical defect and black hole.}
\end{figure}

Both the scalar field and metric functions are finite on the origin, 
\begin{align}\label{ce1}
\phi(0)=2\pi\,,
\qquad 
h(0)=\alpha q^2 e^{-2}\,,
\qquad
f(0)=\alpha q^2 e^2\,,
\end{align} 
(see Fig\ref{f1}). This indicates that when the solutions have a horizon, they are scalar hairy black holes, while when the solutions have no horizon, they are ether scalar hairy conical defects or solitons. We find when $\alpha<0$, (\ref{msoliton}) describes black hole solutions, while when $\alpha>0$, we need to analyze the near-origin behavior of the metric to see whether it is a true soliton or a bare conical singularity. 

With $r\to~0$, the metric goes to
\begin{align}
ds^2=-\frac{\alpha q^2}{e^2}dt^2+\frac{dr^2}{\alpha e^2 q^2}+r^2 d \theta^2\,.
\end{align}
We can see that this spacetime will have conical defect at the origin and thus, lead to a conical singularity unless the condition
\begin{align}
q=\frac{1}{\sqrt{\alpha}e}\,,
\end{align}
is satisfied, which would lead to a true soliton solution with no singularity. This is a reminiscent of the nonrotating BTZ solutions with negative mass, they are also conical singularities except the AdS$_3$ vacuum.

The scalar potential for black holes, conical singularities, and solitons is shown in Fig. \ref{f2}. For black holes when $r\rightarrow 0$ (or $\phi\to 2 \pi$), the scalar potential goes to positive infinity, while for solitons and conical defects, when $r\rightarrow 0$ the scalar potential goes to negative infinity.

\begin{figure}
\centering
\includegraphics[width=0.4 \textwidth]{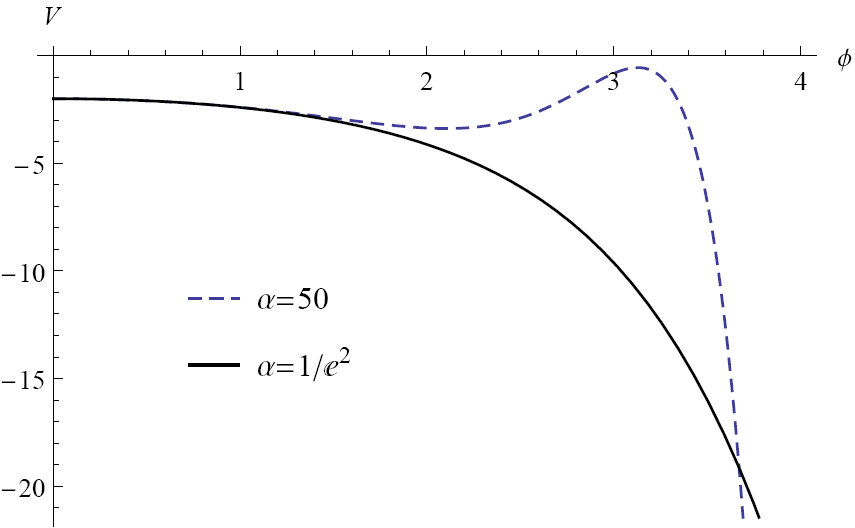}
\hfill
\includegraphics[width=0.4 \textwidth]{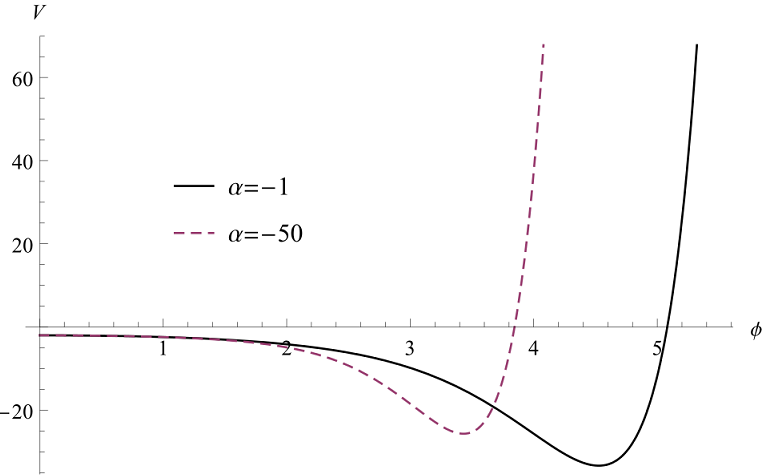}
\caption{\label{f2} Setting $g^2=q=1$, the upper figure shows the scalar potential of a soliton and a conical defect with $\alpha=50$, while the lower figure shows the scalar potential of tow black holes with $\alpha=-1,-50$.}
\end{figure}

The asymptotic behavior of our metric functions is
\begin{align}
&h(r)=g^2r^2+\alpha q^2-\frac{4\alpha  q^3}{3r}+\cdots\,,
\\
&f(r)=g^2r^2+4g^2 q r+(\alpha +4 g^2 )q^2+\cdots\,,
\end{align}
from which we can read $\kappa=\alpha q^2$, so we get
\begin{equation}\label{mass3}
M=2\pi\,\left(\frac{4}{3\ell^2}-\alpha\right)\,q^2\,.
\end{equation}
When $q\rightarrow\,0$, our solutions reduce to the nonrotating massless BTZ solution.

Assume $\alpha<0$ and there is a horizon $r_0=k q$, with $k$ satisfying
\begin{align}
(1-\frac{\alpha}{g^2})k^2+\frac{\alpha}{g^2} e^{-\frac{2}{1+k}}(1+k)^2=0\,.
\end{align}
This equation has one and only one positive solution for $k$ when $\alpha<0$ and indicates $k$ is a function of $\alpha/g^2$. For example when $\alpha/g^2=-1$, we can numerically calculate out $k\approx~0.615$, which indicates $r_0\approx 0.615 ~q$. Taking the small $|\alpha|$ limit, we have
\begin{align}
\alpha\rightarrow 0^{-} ~~~~\Rightarrow~~~~ k\,\rightarrow\,0,
 \end{align} 
while taking the large $|\alpha|$ limit, we have
\begin{align}
\alpha\rightarrow -\infty ~~~~\Rightarrow~~~~k\,\rightarrow\,\sqrt{|\alpha|}\,.
\end{align}
The Hawking temperature $T$ and entropy $S$ are given by
\begin{align}
T=-\frac{\alpha q}{2 k\pi}\,,
\qquad
S=\frac{\pi k q}{2 G}=8\pi^2 k q\,.
\end{align}
It can be easily checked that the thermodynamic first law (\ref{FL2}) is satisfied.

We can also study the thermodynamics of conical defects by Wald's formula. We substitute our scalar field and metric functions into (\ref{wfs}) and get
\begin{subequations}
\begin{align}\label{0int}
\delta\mathcal{H}_{0^{+}}=&-2\pi \sqrt{\frac{h(0)}{f(0)}}\delta f(0)=-4\pi\alpha q\,\delta q=\Phi'_{S}\delta Q_S
\\
\delta\mathcal{H}_{r_\infty}=&\delta M+\frac{16\pi}{3\ell^2}q\delta q=\delta M+\Phi_{S}\delta Q_S
\end{align}
\end{subequations}
where $M$ is given by (\ref{mass3}). Since there is no horizon, there is no entropy, so we interpret the integration (\ref{0int}) as another scalar charge term $\Phi'_{S}\delta Q_S$ rather than a $T\delta S$ term. The first law given by Wald's formula now becomes
\begin{align}
\delta M=(\Phi_{S}+\Phi'_{S})\delta Q_S=4\pi(\frac{4}{3\ell^2}-\alpha)q\delta q\,.
\end{align}

It may be inappropriate to call this equation a thermodynamic first law, since this is not a thermodynamic system. It describes how the energy of the conical singularity changes with its scalar hair. 

\subsection{Higer dimensions}
In higher dimensions the scale invariance of (\ref{eq1})~({\ref{eq3}) is broken; however,   our strategy is still useful for soliton solutions, We can use the symmetry of the EOMs to determine the large $\phi_1$ behavior of the boundary conditions $\phi_2(\phi_1)$. In this situation, we can rescale the EOMs with a big $c$, which can make $\frac{6-2d}{c^2}$ almost vanish, so the rescaled EOMs have approximate scale invariance. This determines the boundary conditions should look like (\ref{exp}) when $\phi_1$ goes to infinity. Some people have already used this method to study the large $\phi_1$ behavior of the boundary condition to study the stability of designer gravity~\cite{Faulkner:2010fh}.

\begin{widetext}
\begin{align}
\label{eq21}
&\frac{\phi'^2}{2}-\frac{3(d-2)h'^2}{4h^2}+\frac{(d-2)h''}{2h}=0\,,
\\
\label{eq22}
&\frac{(d-3)}{r^2}-\frac{(d-3)}{r^2\,f}-\frac{(d-4)f'}{2r\,f}+\frac{(d-2)h'}{2r\,h}-\frac{(d-2)f'\,h'}{4f\,h}-\frac{f''}{2f}=0\,,
\\
\label{eq23}
&\frac{(d-3)(1-f)}{r^2\,h}-\frac{d \, f'}{2rh}-\frac{3(d-2)f\,h'}{2r\,h^2}-\frac{(d+2)f'\,h'}{4h^2}-\frac{(d-4)f\,h'^2}{2h^3}-\frac{f''}{2h}-\frac{f\,h''}{h^2}=\frac{2}{d-2}V(\phi)\,.
\end{align}
\end{widetext}

\section{Constructing exact solutions with spherical metrics with a conformal factor}\label{s3}
\subsection{The strategy}
In this section we apply our strategy to a spherical metric with a conformal factor
\begin{equation}\label{M2}
ds^2=h(r)\left(-f(r) dt^2+\frac{1}{f(r)}dr^2+r^2\,d\Omega_{d-2}^{2}\right)\,.
\end{equation}
This metric looks very like the metric presented in~\cite{Martinez:2004nb,Kolyvaris:2009pc}; however, the topology in~\cite{Martinez:2004nb,Kolyvaris:2009pc} is $\mathbb{R}^{2}\times \Sigma$, where $\Sigma$ is a two-dimensional manifold of negative constant curvature, while the topology of (\ref{M2}) is $\mathbb{R}^2\times S^{d-2}$. The Einstein EOMs in $d$ dimensions (where $d\geq 3$) are (\ref{eq21})~(\ref{eq23}). The scalar field equation of motion (\ref{seom}) can now be expressed as
\begin{align}\label{seom2}
\frac{(d-2)f\,\phi'}{r\,h}+\frac{(d-2)f\,h'\,\phi'}{2h^2}+\frac{f'\,\phi'}{h}+\frac{f\,\phi''}{h}=\frac{\partial V(\phi)}{\partial \phi}\,
\end{align}
and one can also check that, for our solutions in this section, Eq. (\ref{seom2}) would also be automatically satisfied once the EOMs (\ref{eq21})-(\ref{eq23}) are satisfied.

Like the EOMs (\ref{eq1})-(\ref{eq2}) in the previous section, (\ref{eq21})~(\ref{eq23}) also have a scale invariance under a rescaling operation
\begin{align}
&r=c\rho\,,\qquad &~~~~~~~~~~\phi(r)=\tilde{\phi}(\rho)\,,
\cr
&h(r)=\tilde{h}(\rho)\,, \qquad &f(r)=\tilde{f}(\rho)\,,
\cr
&V(\phi)=\frac{\tilde{V}(\tilde{\phi})}{c^2}\,.\qquad &
\end{align}
This means that the function of our scalar field solution should be preserved under such a rescaling operation. The difference is that, here the scale invariance of the EOMs is preserved in general dimensions ($d\geq 3$), thus our strategy should work for metric (\ref{M2}) in general dimensions. It should be promising to find $d\geq 3$ exact solutions  with a given scalar field $Ansatz$ that is an analytical function of $\frac{\phi_1}{r^{p_1}}$. In the rest of this section we will construct some new solutions with our strategy.

\subsection{The thermodynamics}
Before we list our new solutions, we would like to discuss their thermodynamics in general. We substitute our metric (\ref{M2}) into (\ref{Theta}) and get
\begin{align}\label{Fl3}
&\delta\mathcal{H}_{r}=-\frac{h^{\frac{d}{2}-3} r^{d-3}\omega_{d-2}}{2} \Big [2r f h^2 \phi'  \delta\phi+(d-2)\times
\cr
& \big(  (2 h+r h' )h\delta f- r(f' h+3 f h') \delta h+2r f h  \delta h'\big) \Big ]
\end{align}

Consider a black hole with an event horizon $r_0$ which satisfies $f(r_0)=0$, the integral on the horizon becomes
\begin{align}
\delta\mathcal{H}_{r_0}=-&\frac{(d-2)h^{\frac{d}{2}-2} r^{d-3}\omega_{d-2}}{2}\times
\cr
&\left( 2 h  \delta f+r h' \delta f-r f'  \delta h\right)\Big |_{r=r_0}\,.
\end{align}
Using
\begin{align}
&\delta f\big|_{r=r_0}=-f'(r_0)\delta r_0
\\
&\delta h\big|_{r=r_0}=\delta h(r_0)-h'(r_0)\delta r_0\,,
\end{align}
we get
\begin{align}
\delta\mathcal{H}_{r_0}&=\frac{h^{\frac{d}{2}-3}(r_0) r_0^{d-3}\omega_{d-2}}{2}\times
\cr
&\Big[2h^2(r_0)f'(r_0)\delta r_0+r_0 f'(r_0) h(r_0)\delta h(r_0)\Big]\,.
\end{align}
The Hawking temperature and entropy can be derived from (\ref{M2}), which are
\begin{align}
T&=\frac{f'}{4\pi}\big|_{r=r_0}\,,
\\
S&=\frac{\omega_{d-2}r^{d-2}h^{(d-2)/2}}{4 G}\big|_{r=r_0}\,,
\end{align}
where $G=\frac{1}{16\pi}$. As expected we again get
\begin{align}
\delta\mathcal{H}_{r_0}=T\delta S\,.
\end{align}

Then we consider the integral on the boundary $\delta\mathcal{H}_{\infty}$ in Wald's formula. The large $r$ expansion of the scalar fields in this section are all given by
\begin{align}\label{sae}
\phi(r)=\frac{\phi_1}{r}+\frac{\phi_2}{r^2}+\frac{\phi_3}{r^3}+\cdots\,,
\end{align}
which satisfies (\ref{exp}).
If we confine
\begin{align}
h(\infty)\rightarrow\,1\,,
\end{align}
as we will do in the next two subsections, and substitute the large $r$ expansion of $\phi(r)$ into the EOMs (\ref{eq21}) and (\ref{eq22}), we find that when $d=3$ the asymptotic behaviors of the metric functions are given by
\begin{align}
&h(r)=1-\frac{\phi_1^2}{6r^2}-\frac{\phi_1\phi_2}{3r^3}+\cdots\,,
\\
&f(r)=g^2 r^2+\kappa+\frac{\kappa\phi_1^2}{24 r^2}+\frac{\kappa\phi_1\phi_2}{15r^3}+\cdots\,.
\end{align}
Hence we get
\begin{align}\label{fl31}
\delta\mathcal{H}_{\infty}=-\omega_1\delta \kappa=\delta M\,,
\end{align}
and Wald's formula (\ref{wfb}) gives the black hole first law
\begin{align}
\delta M=T\delta S\,.
\end{align} 

It is interesting that there is no contribution from the scalar charge. This is because the scalar field (\ref{sae}) have $\sigma=0$ after we transform our metric (\ref{M2}) into the form (\ref{M1}) with a coordinate transformation on the boundary, which means the conjugate scalar potential (\ref{Scharge}) vanishes.

When $d=4$, the asymptotic behaviors of the metric functions are given by
\begin{align}
&h(r)=1-\frac{\phi_1^2}{12r^2}-\frac{\phi_1\phi_2}{6r^3}+\cdots\,,
\\
&f(r)=g^2 r^2+1+\frac{\kappa}{r}+\frac{\phi_1^2}{12 r^2}+\frac{(\kappa \phi_1^2 + 2 \phi_1 \phi_2)}{20 r^3}+\cdots\,.
\end{align}
From (\ref{Fl3}) we get
\begin{align}\label{fl32}
\delta\mathcal{H}_{\infty}&=-\omega_2(2\delta\kappa-\frac{2}{3} g^2 \phi_2 \delta\phi_1 + \frac{1}{3} g^2 \phi_1 \delta\phi_2)
\cr
&=\delta\left[-\omega_2(2\kappa+\frac{g^2}{3}\phi_1\phi_2)\right]+g^2\omega_2\phi_2\delta\phi_1
\cr
&=\delta M-\Phi_S\delta Q_S
\end{align}
where we have used the definition of mass and scalar charge proposed in~\cite{Wen:2015uma} again
\begin{align}
M&=-\omega_2(2\kappa+\frac{g^2}{3}\phi_1\phi_2)\,,
\\
\Phi_S&=-g^2\omega_2\phi_2\,,
\\
Q_S&=\phi_1\,.
\end{align}
So Wald's formula (\ref{wfb}) gives the black hole first law 
\begin{align}\label{FL4}
\delta M=T\delta S+\Phi_S \delta Q_S
\end{align}

In the last subsection we confine 
\begin{align}
h(r)\propto\frac{1}{r^2}
\end{align}
when $r\rightarrow\,\infty$, the solutions will go to $\mathbb{R}^{1,1}\times S^{d-2}$ on the boundary instead of AdS (or dS) vacuum. We will discuss the thermodynamic of these solutions explicitly in that subsection.

Remember that the above thermodynamic analyses (\ref{fl31}) and (\ref{fl32}) are based on a given scalar $Ansatz$ which satisfies (\ref{sae}); however, it will not be more difficult to generalize our thermodynamic analyses to other scalar fields.

\subsection{3-d solutions}
The first scalar $Ansatz$ we try is
\begin{align}
\phi(r)=\sqrt{2(\mu^2-1)}\text{Arctanh}\frac{1}{1+r/q}
\end{align}
where $\mu\geq 1$. The asymptotic behavior of $\phi(r)$ is given by
\begin{align}
\phi(r)=\frac{\sqrt{2(\mu^2-1)}q}{r}-\frac{\sqrt{2(\mu^2-1)}q^2}{r^2}+\cdots\,,
\end{align}
which satisfies (\ref{exp}). It is difficult to get exact solutions for a general $\mu$, so we only give three examples with $\mu=2\,,3\,,4$.

When \textbf{$\mu$=2}, the metric functions and corresponding scalar potential $V(\phi)$ are given by
\begin{align}
f(r)&=\frac{(g^2-\alpha)r^{5/2}-\alpha\sqrt{r+2q}(q^2+qr-r^2)}{\sqrt{r}}
\\
h(r)&=\frac{r(r+2q)}{(q+r)^2}
\\
V(\phi)&=\frac{e^{-\sqrt{6}\phi}}{16}\big[g^2-\alpha\left(1+8 e^{3\sqrt{\frac{3}{2}}\phi}+24 e^{\frac{5\phi}{\sqrt{6}}}\right)+
\cr
&\left(18e^{2\sqrt{\frac{2}{3}}\phi}+15e^{4\sqrt{\frac{2}{3}}\phi}\right)(\alpha-g^2)\big]\,.
\end{align}

There are black holes in the space of solutions, for example, the solution with $\alpha= g^2$ is quite simple and has an event horizon at $r_0=\frac{1}{2}(q+\sqrt{5}q)$. When $g^2$ is positive (negative or zero), the black hole is asymptotic AdS (dS or flat). After some calculations we find Wald's formula (\ref{wfb}) gives
\begin{align}
\delta\mathcal{H}_{\infty}=\delta\mathcal{H}_{r_0}=5\omega_1\, g^2 q\,\delta q\,,
\end{align}
which indicates $\delta M=T\delta S$.

For the next two cases, we will only list the general solutions. Anyone interested in the black holes can search the parameter space of the solutions.

When \textbf{$\mu$=3}, the metric functions and corresponding scalar potential $V(\phi)$ are given by
\begin{align}
f(r)=&g^2 r^2+\frac{\alpha q(4q^2+6qr+3r^2)}{3r}+\frac{\alpha r^2}{2}\log \frac{r}{2q+r}
\\
h(r)=&\frac{9r^2(2q+r)^2}{(4q^2+6qr+3r^2)^2}
\\
V(\phi)=&\frac{e^{-2\phi}}{18}\big[\alpha+e^{3\phi}(\alpha(3\phi-5)-12g^2)
\cr
&~~~~~~~~+2e^{\frac{3\phi}{2}}(\alpha(3\phi+2)-12g^2)\big]
\end{align}

When \textbf{$\mu$=4}, the metric functions and corresponding scalar potential $V(\phi)$ are given by
\begin{align}
f(r)&=r^2(\alpha+ g^2) -\frac{\alpha \left(2 q^4+4 q^3 r+3 q^2 r^2+q r^3+r^4\right)}{ \sqrt{r^3 (2 q+r)}}
\\
h(r)&=\frac{64q^2r^3(2q+r)^3}{(r^4-(2q+r)^4)^2}
\\
V(\phi)&=\frac{e^{-3\varphi}}{32}\big[40\alpha+24\alpha e^{8\varphi}-e^{\varphi}(\alpha+g^2)\times
\cr
&~~~~~~~~~~~~~~(45+19 \cosh[8\varphi]+16\sinh[8\varphi])\big]
\end{align}
where we have defined $\varphi=\frac{\phi}{\sqrt{30}}$.

\subsection{4-d solutions}
The scalar field $Ansatz$ we choose is
\begin{align}
\phi(r)=\sqrt{3}\log(1+\frac{q}{r})\,,
\end{align}
whose large $r$ expansion satisfies (\ref{exp}) too. The metric functions and corresponding scalar potential $V(\phi)$ are given by
\begin{align}
f(r)=&\frac{\left(4 \left(g^2 r^4+r^2\right)+q^2+4 q r\right)}{4r^2}+\frac{ \sqrt{3}}{2} \alpha r^2 \log \left(\frac{r}{q+r}\right)
\cr
&-\frac{\sqrt{3} \alpha q (q+2 r) \left(q^2+2 q r-2 r^2\right)}{8 r^2}
\\
h(r)=&\frac{4r(q+r)}{(q+2r)^2}
\\
V(\phi)=&(3 \alpha \phi-6 g^2)\cosh [\frac{\phi}{\sqrt{3}}] 
\cr
&-\frac{\sqrt{3} \alpha}{4}  \left(9 \sinh [\frac{\phi}{\sqrt{3}}]+\sinh [\sqrt{3} \phi]\right)\,.
\end{align}
When $\alpha=0$ and $g^2<0$, the solution is an asymptotic de Sitter black hole, with an event horizon 
\begin{align}
r_0=\frac{1+\sqrt{1+2 |g|q}}{2 |g|}\,.
\end{align}
We can also check that the Wald's formula is satisfied by calculating out the integrals $\delta\mathcal{H}_{\infty}$ and $\delta\mathcal{H}_{r_0}$, which are
\begin{align}
\delta\mathcal{H}_{\infty}=\delta\mathcal{H}_{r_0}=-2 \omega_2\delta q
\end{align}

Another $Ansatz$ we consider in four-dimensional spacetime is
\begin{align}
\phi(r)=\sqrt{15}\log(1+\frac{2q}{r})\,.
\end{align}
The metric functions and corresponding scalar potential $V(\phi)$ are given by
\begin{align}
f(r)=&g^2 r^2+\frac{(q+r)^2 \left(2 q^2+2 q r+r^2\right)^2}{r^4 (2 q+r)^2}
\cr
&+\frac{16 q^3 (\alpha-g^2) (q+r)^3 \left(q^2+q r+r^2\right)}{r^4 (2 q+r)^2}
\\
h(r)=&\frac{64 q^2 r^3 (2 q+r)^3}{\left(r^4-(2 q+r)^4\right)^2}
\\
V(\phi)=&6\alpha \sinh ^5[\frac{\phi}{\sqrt{15}}]-6 g^2 \cosh ^5[\frac{\phi}{\sqrt{15}}]
\end{align}

When $\alpha=0$ and $g^2<0$, the solution is also an asymptotic de Sitter black hole with an event horizon
\begin{align}
r_0=\frac{\sqrt{1+4 |g|^2 q^2}+\sqrt{1+4 |g| q \left(\sqrt{4 |g|^2 q^2+1}+|g| q\right)}}{2 |g|}\,.
\end{align}
We can also check that Wald's formula is satisfied
\begin{align}
\delta\mathcal{H}_{\infty}=\delta\mathcal{H}_{r_0}=-4\,\omega_{2} \left(24 |g|^2 q^2+1\right) \delta q\,.
\end{align}

\subsection{Solutions in general dimensions}\label{s34}
We also find some solutions in general $d$ dimensions. Consider a simple scalar field,
\begin{align}
\phi(r)=\frac{\sqrt{2(d-2)}q}{r}\,,
\end{align}
and the metric functions and scalar potential $V(\phi)$ are given by
\begin{align}
f(r)=&\mu r^2 e^{\frac{(d-2) q}{r}}+\frac{2 (d-3) r}{(d-2) q}-
\cr
&\frac{r^2 \left(d\left(\alpha+2\right)-2 \left(\alpha+3\right)\right) \left(e^{\frac{(d-2) q}{r}}-1\right)}{(d-2)^2 q^2}
\\
h(r)=&\frac{\beta e^{-\frac{2 q}{r}}}{r^2}
\\
V(\phi)=&\frac{e^{\frac{\sqrt{2} \phi}{\sqrt{d-2}}}}{\beta} \big[2 \alpha-d \left(\alpha+\sqrt{2} \sqrt{d-2} \phi+5\right)
\cr
&~~~~~~~~~~+d^2+3 \sqrt{2} \sqrt{d-2} \phi+6\big]
\end{align}
with $\mu,\alpha,\beta,q$ as arbitrary constants. We can see that these solutions have two independent physical integration constants $q$ and $\mu$, which do not appear in the Lagrangian.

When $r\rightarrow \infty$, the metric goes asymptotically to
\begin{align}\label{asymptotic solution}
ds^2=\beta\left(-\mu dt^2+\frac{dr^2}{\mu r^4}+ d \Omega_{d-2}^2\right)\,,
\end{align}
Making a coordinate transformation $r=\frac{1}{\rho}$, the asymptotic metric becomes
\begin{align}
ds^2=\beta\left( -\mu dt^2+\frac{d\rho^2}{\mu}+d\Omega_{d-2}^2\right)
\end{align}
which is just a $R^{1,1}\times S^{d-2}$ when $\beta>0$.

To study the thermodynamics, we calculate out $\delta\mathcal{H}_{r}$
\begin{align}
\delta\mathcal{H}_{r}=&-(d-2)\beta^{\frac{d-2}{2}}(q\delta \mu+2 \mu\delta q)
\end{align}
which is independent of $r$.
\begin{align}
\delta\mathcal{H}_{\infty}=&\delta[-(d-2)\beta^{\frac{d-2}{2}}q \mu]-(d-2)\beta^{\frac{d-2}{2}}\mu\delta q
\cr
=&\delta M-\Phi_S\delta Q_S
\end{align}
where, according to~\cite{Wen:2015uma}, we have defined 
\begin{align}
M=&-(d-2)\beta^{\frac{d-2}{2}}q \mu \,,\qquad \Phi_S=\frac{\sqrt{d-2}}{2}\beta^{\frac{d-2}{2}}\mu\,,
\cr
Q_S=&\sqrt{2(d-2)}q\,.
\end{align}

So for black holes, Wald's formula gives
\begin{align}
\delta M=T\delta S+\Phi_S\delta Q_S\,.
\end{align}

In three dimensions, equation $f(r_0)=0$ has an analytical solution, which is given by
\begin{align}
r_0= -\frac{q}{\log \left(1-\frac{\mu q^2}{\alpha}\right)}\,.
\end{align}
As expected, we find
\begin{align}
\delta\mathcal{H}_{r_0}=T\delta S=-\beta^{\frac{1}{2}}(q\delta \mu+2 \mu\delta q)=\delta\mathcal{H}_{\infty}\,.
\end{align}

In higher dimensions, it is hard to solve $f(r_0)=0$ analytically, so we need to consider more specific solutions. For example, if we set $\beta=\mu=\alpha=1$, $d=4$, and $q=-1$, the solution have a horizon at $r_0=1$.

\section{Conclusion and future prospects }

Our strategy should also be useful for metrics with other topologies (for example the planar metric) and Einstein-scalar gravities with other kinds of couplings (for example the nonminimal coupling) if the EOMs have a scale invariance. For example, in~\cite{Feng:2013tza} we used a metric $Ansatz$ inspired by the construction of black $p$-branes,
\begin{equation}\label{m1}
ds^2=-\frac{f(r)}{h(r)}dt^2+h(r)^{\frac{1}{d-3}}\left(\frac{dr^2}{f(r)}+r^2 d\Omega_{d-2}^{2}\right)
\end{equation}
where again, $d$ represents the spacetime dimension, and $d\Omega_{d-2}^{2}$ is the metric of unit $d-2$ sphere. In general $d\geq 4$ dimensions , the EOMs also have a scale invariance under the following rescaling operation
\begin{align}
&r=c\rho\,,\qquad &\phi(r)=\tilde{\phi}(\rho)\,,~~~~~~~~~~~~~~~
\nonumber\\
&h(r)=\tilde{h}(\rho)\,,\qquad &f(r)-1=c^2(\tilde{f}(\rho)-1)\,,
\cr
&V(\phi)=\tilde{V}(\tilde{\phi})
\end{align}
The solutions given in~\cite{Feng:2013tza} also satisfy (\ref{exp}) and have only one physical integration constant, which describes the scalar hair, and some of the the corresponding scalar potential can be expressed in terms of a superpotential. It would be interesting to find out whether the scalar potentials we found in this paper have supersymmetry.

The main purpose of this paper is to give an efficient strategy to find out what kind of metric $Ansatz$es may admit simple solutions, how to construct the exact solutions, and what are the most general theories which admit the scalar field $Ansatz$ as a solution. We use the symmetry of the EOMs to choose the right promising $Ansatz$ for $\phi(r)$ and then derive the other metric functions and scalar potential. Our strategy is useless if we start from a specific theory, however, if this specific theory does admit simple exact solutions, it should be promising to find them out by searching the scalar fields with our strategy. For example, we reproduced the HMTZ black holes in Sec. \ref{s2}. So far, all the known scalar hairy exact solutions have a scalar field that satisfies (\ref{exp}), which means they can be reproduced and some of them can be generalized to a larger class by our strategy.  

However when we make the scalar field $Ansatz$ satisfy (\ref{exp}), we drop the other integration constant, for example $r_0$, so very likely there is only one physical integration constant in our solutions, which describes the scalar hair, and a boundary condition $\phi_2(\phi_1)$ is automatically given. When we turn off this integration constant, we just get the massless, rather than the massive, static BTZ solutions. To get static asymptotic AdS solutions with two integration constants, the most direct way is to start from a more general scalar field $Ansatz$ whose asymptotic behavior satisfies (\ref{phii}). However this would very likely produce difficulties for finding analytic solutions, as the scalar field $Ansatz$ gets more complicated. We still need some luck to get such exact solutions.

\begin{acknowledgements}
I am grateful to Hai-Shan Liu, Hong Lu,Rob Myers  and especially Wei Song for helpful comments and reading the paper. I would also like to thank Chuan-jie Zhu, the Beijing Municipal Education Committee and the Perimeter Institute Visiting Graduate Fellows program for financial support. Research at Perimeter Institute is supported by the Government of Canada and by the Province of Ontario through the Ministry of Research and Innovation. This work is also supported in part by NSFC 34112027.
\end{acknowledgements}

%\bibliographystyle{unsrt}
%\bibliography{/home/thiago/bibtex/articles,/home/thiago/bibtex/books}

\end{document}